# Facilitating method development for reverse fill/flush flow modulation by using tuneable auxiliary pressure source instead of a fixed bleed capillary

Aleksandra Lelevic[a,*],

a. TotalEnergies, Centre de Recherche de Solaize, Chem. du Canal, 69360 Solaize, France

* Authors for correspondence: aleksandra.lelevic@totalenergies.com

**Abbreviations**

1D – One dimensional

$^1$D – First dimension

$^2$D – Second dimension

CFT – Capillary flow technology

GC×GC – Comprehensive two-dimensional gas chromatography

EPC – Electronic pneumatics control

PCM – Pressure control module

RFF – Reverse fill/flush flow modulation

# Abstract

The conventional reverse fill/flush flow modulation for comprehensive two-dimensional gas chromatography requires a bleed capillary column to be connected to the outlet of the modulator channel. The purpose of this capillary, that does not contain the stationary phase, is to provide a pressure resistance to the modulator channel flow. In this way, the desired modulator flow can be achieved, and channel over-filling can be avoided. Normally, the length and the internal diameter of the bleed capillary are chosen so as to obtain the modulator flow that is close to the flow of the first separation column. Thus, for any chosen set of chromatographic conditions, the required dimensions of the bleed capillary can be completely different, making the GC×GC method development tedious and generating additional costs in consumables and analyst time. In this work a tuneable pressure source generating a suitable backpressure was used instead of the fixed bleed capillary which has the advantage of the possibility to freely adapt the pressure resistance and generate the required modulator channel flow for any conditions. This set-up has been evaluated and compared in terms of the impact on the modulation performance to the set-up involving a fixed bleed capillary demonstrating comparable performance.

1. Introduction

Modulator is a key element of a comprehensive two-dimensional gas chromatography (GC×GC) system. Its function is to successively sample the effluent from the first-dimension column and then introduce it in a focused band in the second separation column. Modulators can be classified into three types: thermal, valve-based and flow modulators [1–5].

Valve-based and flow modulators have gained in importance over the last years due to their low cost, ease of use and amenability to samples containing compounds with a wide range of boiling points. Flow modulation for GC×GC was first proposed by Bruckner et al. [6] who used a diaphragm valve to divert the $^1$D GC effluent to the $^2$D column. Seeley et al. [7] modified the initial design and devised a system that involved a sample loop with a diaphragm valve. Bueno and Seeley [8] then further refined the approach by eliminating the diaphragm valve and thus allowing a 100% mass transfer.

Flow modulators differ from valve-based ones as in these systems the two column flows are coupled. Seeley et al. [9] in 2006 have introduced a first modulator of this type that involved a Deans switch in which 100% of $^1$D effluent was transferred to the $^2$D column. This design served as a basis for the first commercialization of flow modulators: a Capillary Flow Technology (CFT) by Agilent Technologies. Flow modulators have been compared to cryo modulators. Semard et al. [10], found that the differential flow modulation can result in a similar peak capacity as cryo modulation in the optimized conditions.

CFT modulator with a fixed integrated modulator channel comes in two types: forward fill/flush (FFF) and reverse fill/flush (RFF) modulation. Both of these systems operate by repeating in cycles the stages of filling and flushing of the modulator channel, which are driven by the periodic switching of the three-way solenoid valve that receives a constant supply of the carrier gas. The schematic representation of both systems is provided in Figure S1 in the Supporting information. The FFF modulation system was employed in many works with good performance [10–18]. The RFF modulation system differs from the FFF by the fact the modulator channel is flushed in the inverse direction to the direction of the channel filling. It has been demonstrated that such an approach aids to obtain narrower peaks due to the effect of the analyte band contraction [19]. Solution for the CFT RFF flow modulator with an adjustable modulator channel accumulation loop also exists and involves combining a 2-way non-purged splitter and 2-way purged splitter CFT plates [19,20]. Another commercial option of the RFF system is the SepSolve Analytical INSIGHT modulator. It is a single plate modulation system with an adjustable accumulation loop. Both types of modulator technologies have been extensively applied across various applications. Following references involve the use of CFT reverse fill/flush modulation [19–27], while [28–34] employ INSIGHT reverse fill/flush modulator.

In the RFF system during the "fill" stage of the modulation, the $^2$D inlet pressure is no longer applied at the outlet of the modulator channel (Figure S1C). This is why a bleed capillary with a specific length and internal diameter is installed at the outlet of the modulator channel so as to create a suitable backpressure and control the flow inside the modulator channel. In this way, during the fill stage of the modulation, channel over-filling will be avoided. The required dimensions of the bleed capillary will be completely different for different sets of chromatographic conditions. This can make the method development on such a system quite tedious since due to testing of different separation conditions reinstallation of the different bleed capillary is required. This might inhibit the fine tuning of the separation conditions but also represents additional costs in consumables and analyst time.

As the role of the bleed capillary is nothing more than to generate a pressure resistance to the modulator channel flow, in this work for this purpose an auxiliary pressure source was coupled with an CFT RFF that enabled to freely tune the pressure resistance for generating the desired modulator channel flow. This has enabled to avoid changing of the bleed capillary for different chromatographic conditions and has enabled easier method development. Such an approach can be interesting to incorporate in commercial RFF systems as every day increase in the use of the GC×GC in research but also in industry can benefit from the approaches that can allow easier and faster method development. The proposed set-up has been tested for a test mixture containing different hydrocarbons and a gas oil sample and its performance in terms of obtained chromatographic separation and quality of modulation has been compared to the system with a fixed bleed capillary.

## 2. Materials and methods

A gas oil reference sample (Part No 00.02.719) and a gravimetric blend of different hydrocarbons (Part No 00.02.719) were obtained from Analytical Controls by PAC.

For GC×GC analysis 7890A gas chromatograph was employed, equipped with an G4573A reverse fill/flush CFT differential flow modulator (0.167 m × 0.535 mm ID), both by Agilent Technologies. Employed column set involved: $^1$D column DB-1 (100% dimethylpolysiloxane;

20 m × 0.1 mm ID × 0.4 μm; Agilent Technologies) and $^2$D column ZB-50 (50% phenyl polysilphenylene-siloxane; 9.5 m × 0.25 mm ID × 0.1 μm; Phenomenex). The lengths of the columns were determined by injecting MeOH and determining columns hold time. $^2$D separation column was connected to an FID detector.

Two configurations of the system were tested in terms of restrictor capillary: i) a deactivated silica restrictor capillary (1.3 m × 0.05 mm ID; Agilent Technologies), and ii) a system involving an auxiliary pressure source. In the latter case, a purged two-way splitter (G3180-61500; Agilent Technologies) receiving a supply of hydrogen gas from an auxiliary pressure control module (Aux EPC) was connected to the modulator channel of the CFT RFF plate with a help of a deactivated silica restrictor capillary (0.5 m × 0.1 mm ID; Agilent Technologies). A first flow splitting port of the two-way splitter was connected to an FID detector with an identical deactivated silica capillary (0.5 m × 0.1 mm ID; Agilent Technologies). Second port of the two-way splitter was simply plugged with a nut and ferrule with a stainless-steel wire. In both set-ups bleed capillary exit was connected to an FID detector. A scheme of the system is shown in Figure 1 with directions of flows of hydrogen gas in the two modulation stages "fill" and "flush" indicated with arrows. Figure S2 in the Supporting information shows the column configuration entered into Agilent OpenLab software.

Carrier gas was hydrogen. 1 μL injection with a split ratio of 300:1 was performed on an Agilent split/splitless inlet equipped with a single taper liner with glass wool. Injection port was heated to 325 °C. Flow in the first dimension was 0.2 mL/min and 20 mL/min in the second dimension. Modulation period was 5 s with 0.2 s modulator injection time. Oven temperature program was: 40°C (1min) – 41°C (0.2°C/min); 41°C – 42°C (0.5°C/min); 42°C – 43°C (1°C/min); 43°C – 320°C (2°C/min) with 147.5 min analysis duration. Slow ramp in the beginning of the run was applied for better separation of light hydrocarbons. FID conditions were as follows: temperature 325 °C, air flow 400 mL/min, hydrogen 20 mL/min, make-up gas (nitrogen) 40 mL/min, acquisition frequency 100 Hz. For the secondary FID connected to the fixed bleed capillary same conditions were applied with hydrogen flow of 40 mL/min. In the case of an auxiliary pressure source, total hydrogen flow was 40 mL/min (auxiliary $H_2$ flow + FID detector $H_2$ flow).

OpenLab CDS (Agilent Technologies) software v.2.4 was used for data acquisition. GC Image GC×GC Software v.2021 was used for visualization and integration of the GC×GC data.

## 3. Results and discussion

The scheme of the RFF system with a fixed bleed capillary is shown in Figure S1 in the Supporting information. During the "fill" stage of the modulation, the $^1$D column effluent fills the modulator channel (Figure S1 C). In the "flush" stage the content of the modulator channel is sent in a quick pulse towards the $^2$D column (Figure S1 D). If a fixed length bleed capillary is employed, a required length of tubing with a given diameter to reach required RFF modulator channel flow can be calculated by using pneumatic model for the reversed-flow differential flow modulator devised by Giardina et al. [25]. This model employs the Poiseuille's Law and the law of the conservation of mass in order to calculate volumetric flows and average velocities in all segments of the fluid circuit representing the system. In this model, a 20% higher flow in the modulator channel compared to the $^1$D column flow was recommended in order to prevent flow splitting between the modulator channel and $^2$D column during the "fill" modulation stage. Table 1 shows calculated bleed capillary lengths, according to Giardina et al. model, required in order to achieve indicated modulator channel flows. It can be seen that for most flow condition changes a new restrictor capillary must be installed.

In order to circumvent this issue, we have devised a set-up in which we have employed an auxiliary pressure source to tune the pressure resistance (backpressure) required for generating the desired modulator channel flow. The scheme of the set-up is provided in Figure 1 with directions of flows of hydrogen gas in the two modulation stages indicated with arrows. The purpose of an auxiliary pressure source is to play the role of a bleed capillary with the advantage that its pressure can be freely adjusted without the need to install a new capillary. Corresponding fluid circuit for the "fill" stage of the modulation process is shown in Figure 2.

For the first set of conditions in Table 1, $^1$D flow 0.2 mL/min and $^2$D flow 20 mL/min, we can calculate the required settings of the auxiliary pressure controller in order to achieve a modulator channel flow of 0.24 mL/min. In the fluid circuit representing the system shown in Figure 2, due to the law of the conservation of mass, normalized volumetric flows at the outlet of the modulator channel and at the outlet of restrictor capillary 1 must be equal:

$$F_3 = F_4 \quad (1)$$

According to Poiseuille's Law, modulator channel normalized volumetric flow can then be expressed by a following equation:

$$F_3 = \left(\frac{\pi r_{mod}^4}{16\eta L_{mod}}\right)\left(\frac{P_2^2 - P_4^2}{P_4}\right)\left(\frac{T_{ref}}{T}\right)\left(\frac{P_4}{P_{ref}}\right) \quad (2)$$

Where $F_3$ is the normalized volumetric flow (m³/s) at the outlet of the modulator channel, $r_{mod}$ is the modulator channel radius (m), $\eta$ is the carrier gas viscosity (Pa·s), $L_{mod}$ is the modulator channel length (m), $P_2$ is the modulator channel inlet pressure (Pa), $P_4$ is the modulator channel outlet pressure (Pa), $T$ is the oven temperature (K), $T_{ref}$ is the reference temperature, 25 °C (298 K), $P_{ref}$ is the reference pressure, 14.696 psia (101 325 Pa).

$P_2$ is equal to the ²D column head pressure, in case of 20 mL/min flow and oven temperature of 40°C $P_2 = 39.73102$ psia. Equation 2 permits to calculate the outlet pressure of the modulator channel $P_4$ required to reach the desired normalized volumetric flow of 0.24 mL/min, $P_4 = 39.73085$ psia.

Restrictor capillary 1 flow can then be expressed with the following equation:

$$F_4 = \left(\frac{\pi r_{res1}^4}{16\eta L_{res1}}\right)\left(\frac{P_4^2 - P_5^2}{P_5}\right)\left(\frac{T_{ref}}{T}\right)\left(\frac{P_5}{P_{ref}}\right) \quad (3)$$

Where $F_4$ is the normalized volumetric flow (m³/s) at the outlet of the restrictor 1, $r_{res1}$ is the restrictor radius (m), $L_{res1}$ is the restrictor length (m), $P_4$ is the restrictor inlet pressure (Pa), $P_5$ is the restrictor outlet pressure (Pa).

As the flow at the outlet of the restrictor 1 is equal to the modulator channel flow, 0.24 mL/min, and modulator outlet pressure equal to the inlet pressure of the restrictor 1, equation 3 permits to calculate the restrictor 1 outlet pressure $P_5$, in this case 39.3062 psia. As restrictor 1 outlet pressure is equal to the restrictor 2 inlet pressure, this permits to calculate the required flow at the outlet of the restrictor 2 (that will be controlled by the auxiliary pressure source) to reach the modulator channel flow of 0.24 mL/min:

$$F_5 = \left(\frac{\pi r_{res2}^4}{16\eta L_{res2}}\right)\left(\frac{P_5^2 - P_6^2}{P_6}\right)\left(\frac{T_{ref}}{T}\right)\left(\frac{P_6}{P_{ref}}\right) \quad (4)$$

Where $F_5$ is the normalized volumetric flow (m³/s) at the outlet of restrictor 2, $r_{res2}$ is the restrictor radius (m), $L_{res2}$ is the restrictor length (m), $P_5$ is the restrictor inlet pressure (Pa), $P_6$ is the restrictor outlet pressure (Pa) - 14.696 psia in case of an FID detector.

Finally, a flow of 9.503 mL/min at the outlet of the restrictor 2 was calculated according to equation 4. This flow will result in a restrictor 1 flow, and thus modulator channel flow, of 0.24 mL/min. In the same way, for all the other conditions from table 1 a required restrictor 2 flow can be calculated (see Table 2). Excel file permitting to perform calculations is provided in Supporting information.

A gas oil reference sample was analysed by both set-ups, the one involving a fixed bleed capillary and with a set-up involving an auxiliary pressure source. Identical peak distribution was obtained as seen in Figure 3 which shows the obtained chromatograms. This is also testified by very comparable superposition of "unfolded" 1D chromatograms for the same analysis (see Figure S3 in the Supporting information). In both set-ups, no modulator channel over-filling was obtained as monitored by the secondary FID detector and as demonstrated in Figure S4 in the Supporting information which shows a superposition of the secondary FID detector signals acquired during gas oil analysis. Gas oil analysis has demonstrated that the pressure resistance of the auxiliary pressure source is comparable to the pressure resistance of the fixed bleed capillary and that theoretical calculation offer a good prediction for chromatographic performance.

To investigate the modulation performance in more detail in the two cases, a gravimetric test mixture containing different hydrocarbon species was also analysed. Obtained chromatograms are provided in Figure 4. In this case also, highly comparable chromatograms were obtained and a total of 43 hydrocarbon species could have been quantified. Figure S3 in the Supporting information shows the superposition of the "unfolded" 1D chromatograms for the gravimetric test mixture. A zoom into a n-C9 modulated peak is shown in Figure 5, demonstrating modulated peaks with comparable peak width and intensity with a slight shift of the 1D peak retention time that can be attributed to small differences in theoretical and real column lengths.

In order to look further into the modulation performance, 43 peaks were integrated in GC Image software and peak volume, along with $^1$D and $^2$D retention times for all 43 species were compared in the two chromatograms. Results of this comparison were provided in Table 1 in the Supporting information. For peak volumes, relative standard deviations lower than 3% were obtained, $^1$D retention time variation was less than 0.3%, $^2$D retention times varied up to 4%, all this demonstrating very good agreement between the two approaches and testifying that theoretical calculations offer good prediction despite inevitable small deviations between estimated and actual column dimensions.

As the role of the auxiliary pressure source is to freely tune the pressure resistance to the modulator flow, other conditions of $^1$D and $^2$D flows were tested in combination with corresponding auxiliary pressure source flows as listed in Table 2. Figure 6 illustrates n-paraffin modulated peak obtained in four different conditions of $^1$D and $^2$D flows. It can be seen that peaks are properly modulated (Figure 6A) with no signal generated on the secondary FID detector (Figure 6B) testifying once more the feasibility of the approach.

4. Conclusions

In this work a tuneable pressure source was employed for reverse fill/flush modulation instead of the fixed bleed capillary which has been demonstrated to produce comparable result in terms of peak modulation with the advantage of the possibility to freely adapt the pressure resistance (backpressure) and generate required modulator channel flow for any conditions. A required auxiliary pressure source flow which will offer equivalent pressure resistance to the one of the fixed restrictor capillary for the used set-up was calculated by employing a Poiseuille's Law and the law of the conservation of mass. Experimental results have proven to be well in line with theoretical calculations offering comparable performance for the two systems. The approach with an auxiliary pressure source has however permitted to avoid changing of the fixed bleed capillary for any change of other system conditions and has allowed more flexibility but also reduction of costs in consumables and analyst time and thus can be a good solution to consider for future RFF commercial systems.

**List of figures:**

Figure 1 Schematics of the employed reverse fill/flush flow modulation system. (A) In the "fill" stage modulator channel is filled with the $^1$D effluent; B) In the "flush" stage modulator channel content is sent in a quick pulse to the $^2$D column. Aux EPC is employed to tune the pression resistance and obtain the desired modulator channel flow. PCM - pressure control module.

Figure 2 Fluid circuit for the "fill" stage of the modulation representing the employed RFF system. P is the pressure (Pa) and F is the normalized volumetric flow (m$^3$/s). $P_3$ and $P_6$ are outlet pressures equal to the atmospheric pressure 14.696 psia.

Figure 3 Chromatograms obtained for a gas oil sample in a set-up with: (A) a fixed bleed capillary, and (B) an auxiliary pressure source.

Figure 4 Chromatograms obtained for a mixture of hydrocarbons sample in a set-up with: (A) a fixed bleed capillary, and (B) an auxiliary pressure source.

Figure 5 n-C9 modulated peak obtained for the analysis of a mixture of hydrocarbons in a set-up with: a fixed bleed capillary, and an auxiliary pressure source.

Figure 6 n-paraffin modulated peak obtained in a set-up with an auxiliary pressure source for different conditions of $^1$D and $^2$D flows (A) Signal on FID connected to $^2$D column, (B) Signal on FID connected to restrictor capillary outlet.

Table 1 Required bleed capillary lengths to achieve corresponding "fill" stage modulator channel flows for the indicated conditions of $^1$D and $^2$D column flows.

| $^1$D Flow (mL/min) | $^2$D Flow (mL/min) | Mod. channel Flow (mL/min) | Bleed capillary ID (mm) | Bleed capillary length (m) |
|---|---|---|---|---|
| 0.20 | 20 | 0.24 | 0.05 | 1.27 |
| 0.20 | 17 | 0.24 | 0.05 | 1.08 |
| 0.20 | 15 | 0.24 | 0.05 | 0.95 |
| 0.15 | 20 | 0.18 | 0.05 | 1.69 |
| 0.15 | 17 | 0.18 | 0.05 | 1.44 |
| 0.15 | 15 | 0.18 | 0.05 | 1.27 |

Table 2 Required auxiliary flows for the restrictor 2, necessary to achieve corresponding "fill" stage modulator channel flows for the indicated conditions of $^1$D and $^2$D column flows.

| $^1$D Flow (mL/min) | $^2$D Flow (mL/min) | Mod. channel Flow (mL/min) | Aux flow (restrictor 2) (mL/min) |
|---|---|---|---|
| 0.20 | 20 | 0.24 | 9.50 |
| 0.20 | 17 | 0.24 | 8.04 |
| 0.20 | 15 | 0.24 | 7.07 |
| 0.15 | 20 | 0.18 | 9.56 |
| 0.15 | 17 | 0.18 | 8.10 |
| 0.15 | 15 | 0.18 | 7.13 |

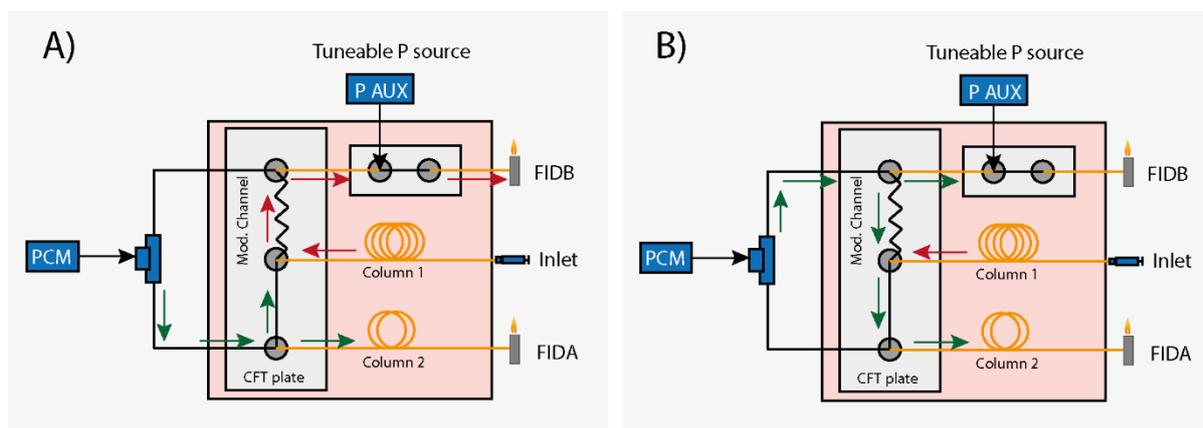

Figure 1 Schematics of the employed reverse fill/flush flow modulation system. (A) In the "fill" stage modulator channel is filled with the $^1$D effluent; B) In the "flush" stage modulator channel content is sent in a quick pulse to the $^2$D column. Aux EPC is employed to tune the pression resistance and obtain the desired modulator channel flow. PCM - pressure control module.

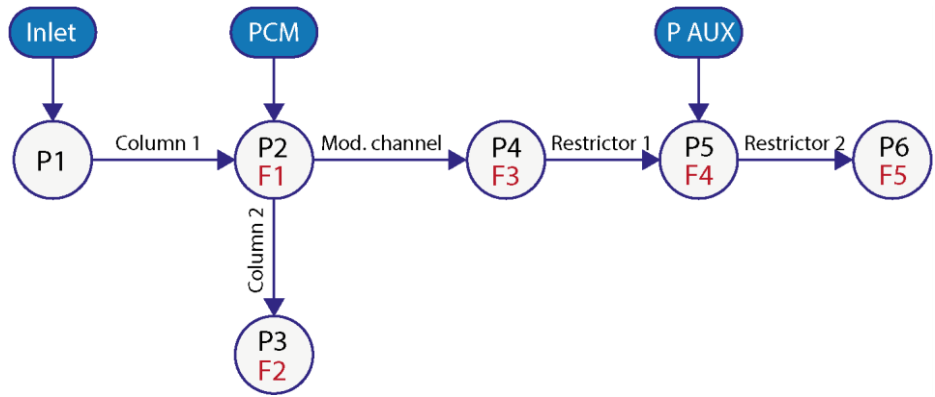

Figure 2 Fluid circuit for the "fill" stage of the modulation representing the employed RFF system. P is the pressure (Pa) and F is the normalized volumetric flow (m$^3$/s). $P_3$ and $P_6$ are outlet pressures equal to the atmospheric pressure 14.696 psia.

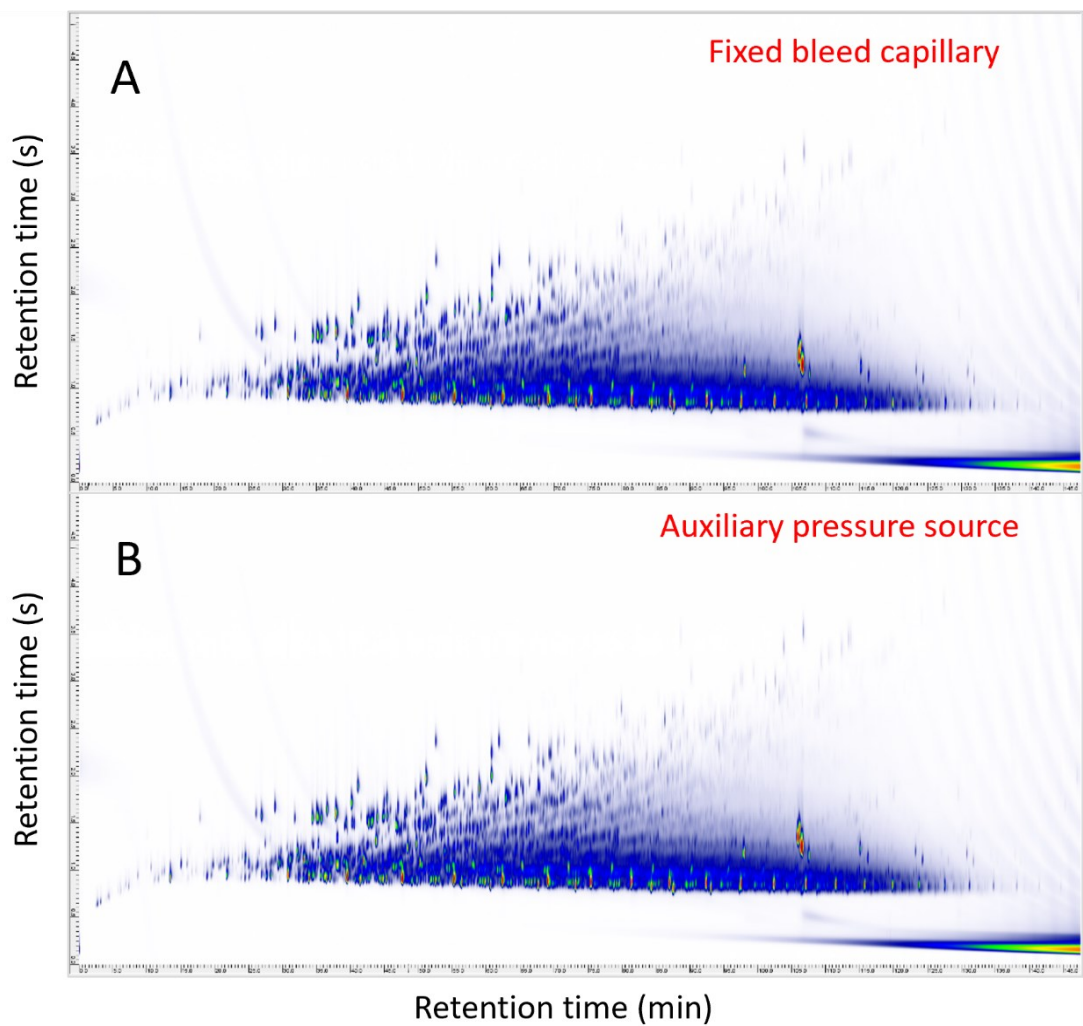

Figure 3 Chromatograms obtained for a gas oil sample in a set-up with: (A) a fixed bleed capillary, and (B) an auxiliary pressure source.

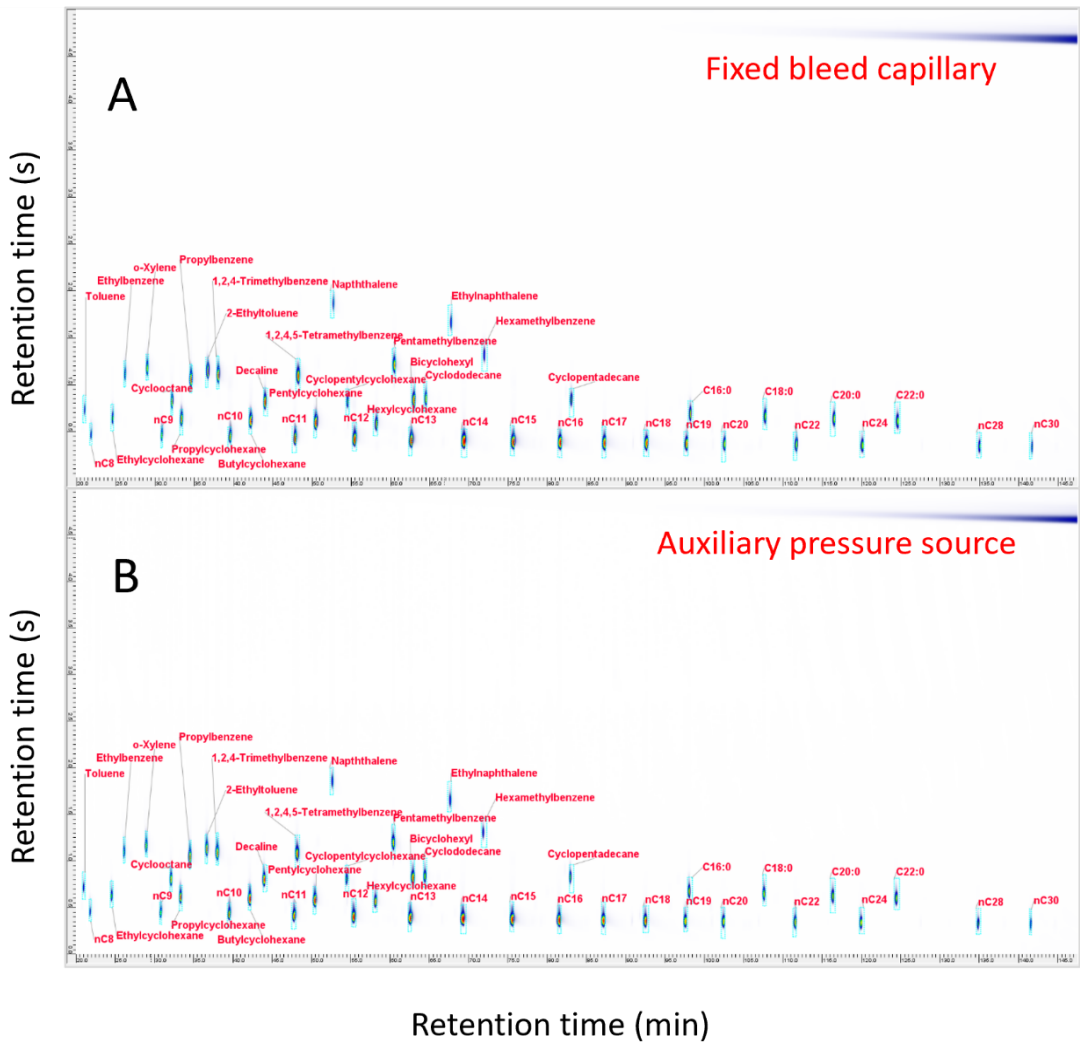

Figure 4 Chromatograms obtained for a mixture of hydrocarbons sample in a set-up with: (A) a fixed bleed capillary, and (B) an auxiliary pressure source.

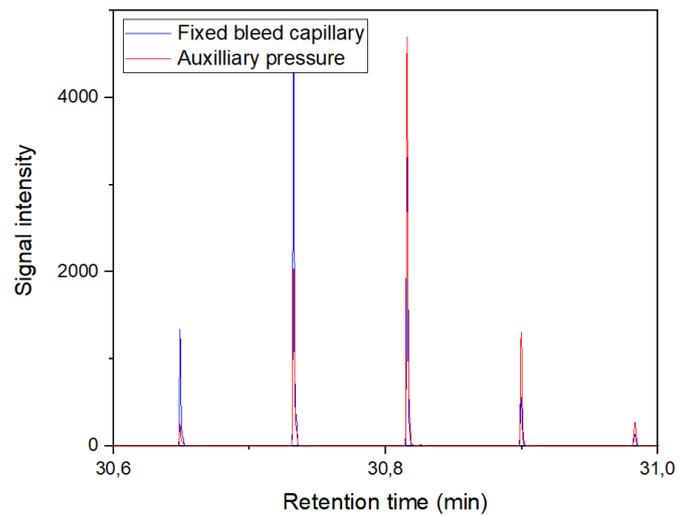

Figure 5 n-C9 modulated peak obtained for the analysis of a mixture of hydrocarbons in a set-up with: a fixed bleed capillary, and an auxiliary pressure source.

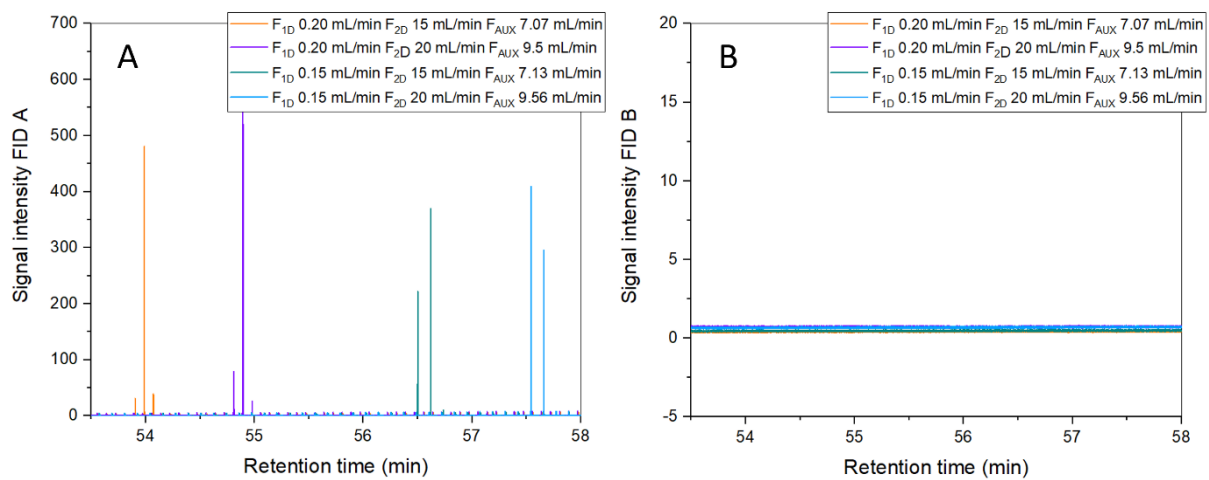

Figure 6 n-paraffin modulated peak obtained in a set-up with an auxiliary pressure source for different conditions of $^1D$ and $^2D$ flows (A) Signal on FID connected to $^2D$ column, (B) Signal on FID connected to restrictor capillary outlet.

# Supporting information

# Facilitating method development for reverse fill/flush flow modulation by using tuneable auxiliary pressure source instead of a fixed bleed capillary

Aleksandra Lelevic[a,*],

a. TotalEnergies, Centre de Recherche de Solaize, Chem. du Canal, 69360 Solaize, France

* Authors for correspondence: aleksandra.lelevic@totalenergies.com

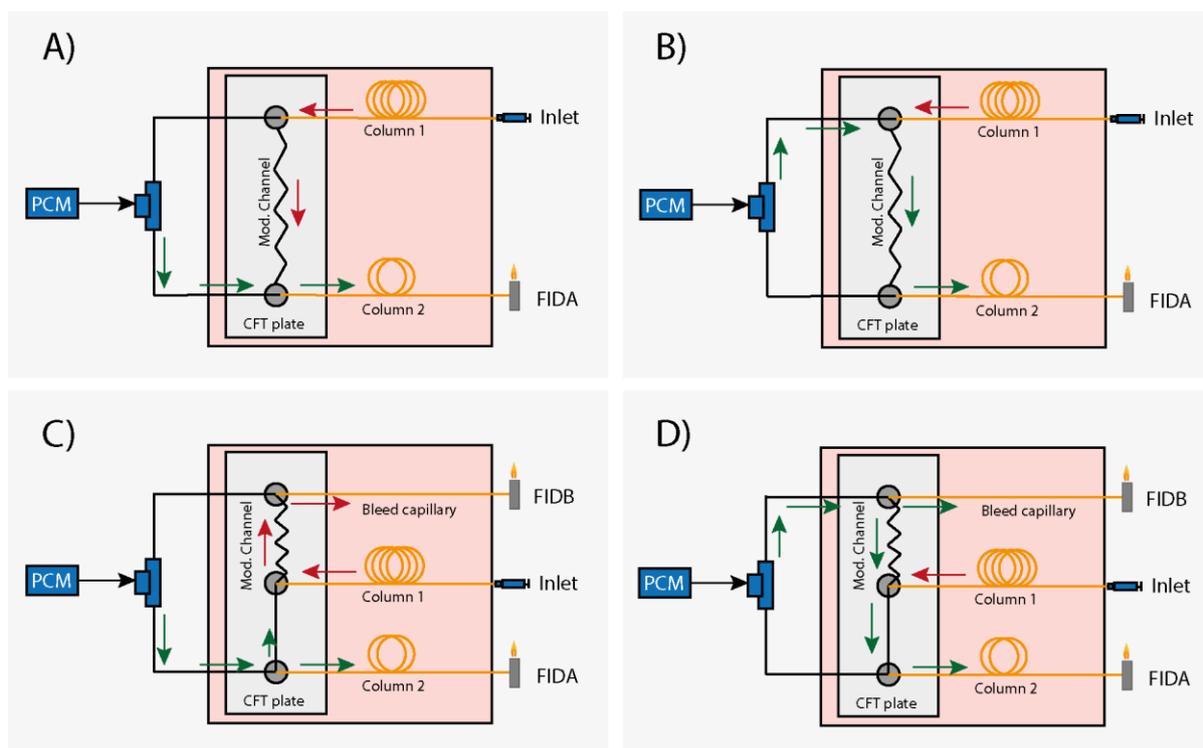

Figure S1 Scheme of the CFT forward fill/flush (FFF) and reverse fill/flush (RFF) modulation systems. A) FFF fill stage, B) FFF flush stage; C) RFF fill stage, D) RFF flush stage.

| | |
|---|---|
| **Column 1 → PCM C** | **0.2 mL/min → PCM C** |
| **PCM C → Column 2 → FID A** | **PCM C → 20 mL/min → FID A** |
| **PCM C → Restrictor 1 → AUX EPC 6** | **PCM C → 0.24 mL/min → AUX EPC 6** |
| **AUX EPC 6 → Restrictor 2 → FID B** | **AUX EPC 6 → 9.5 mL/min → FID B** |

Figure S2 Column configuration settings in the Agilent OpenLab software.

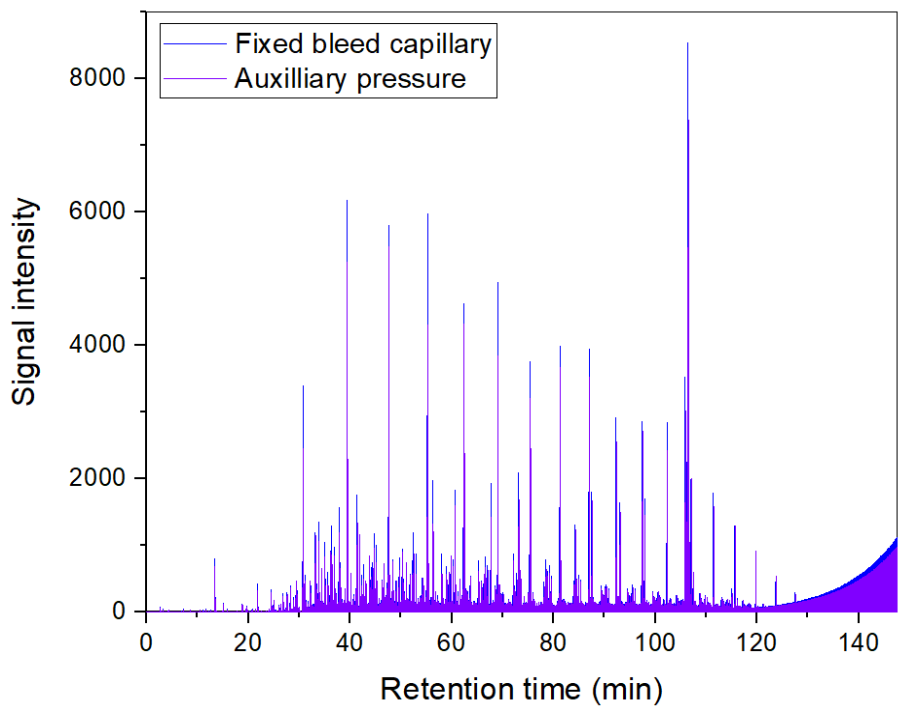

Figure S3 Overlayed gas oil 1D chromatograms obtained for a gas oil sample in a set-up with: a fixed bleed capillary, and an auxiliary pressure source.

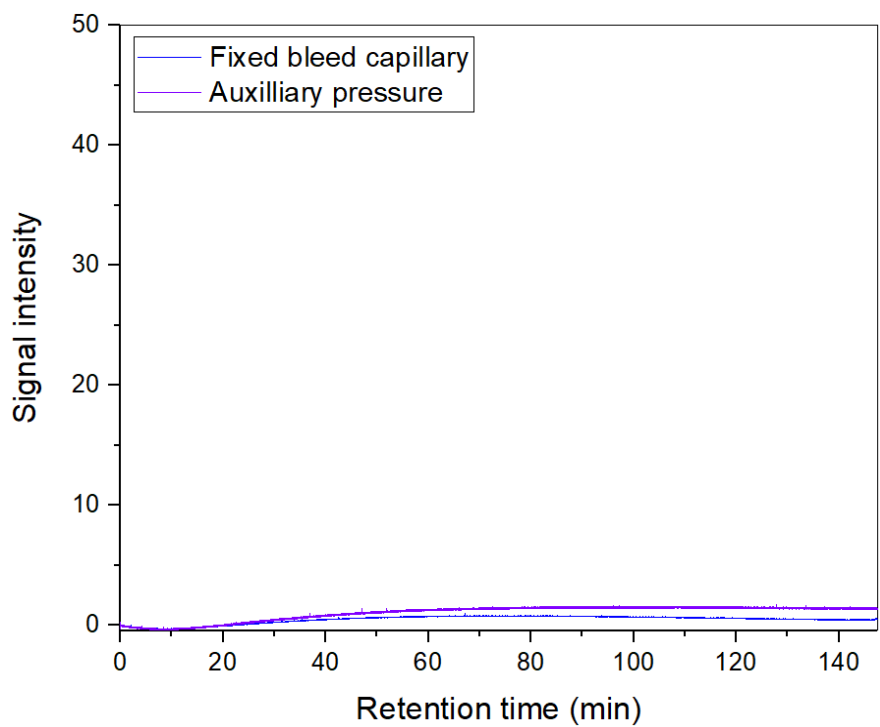

Figure S4 Overlayed gas oil 1D chromatograms obtained for a gas oil sample on the secondary FID in a set-up with: a fixed bleed capillary, and an auxiliary pressure source.

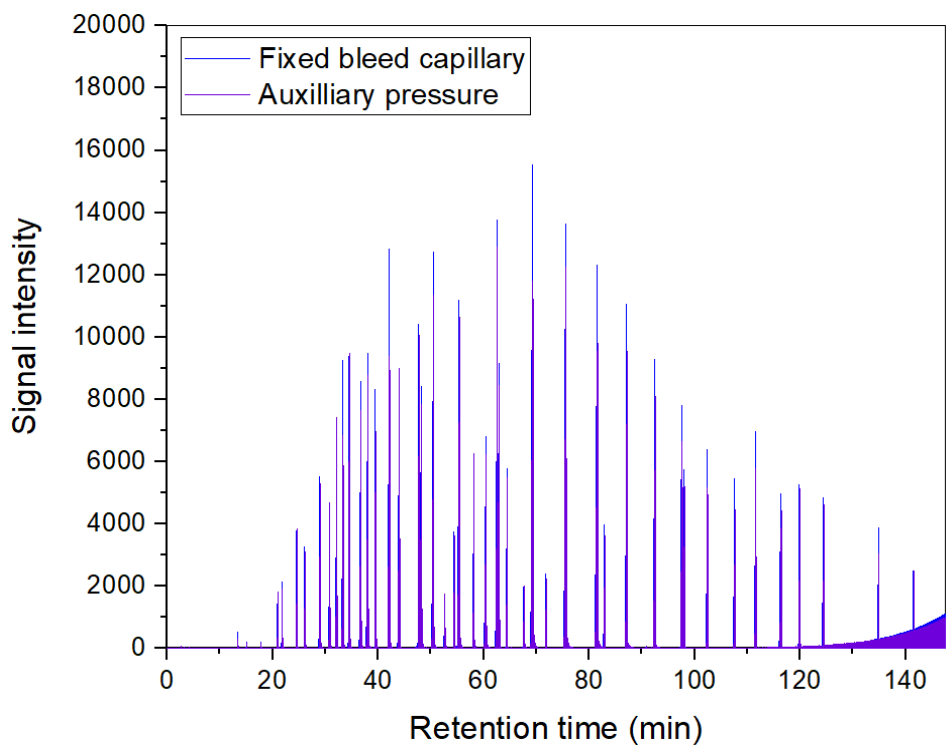

Figure S5 Overlayed gas oil 1D chromatograms obtained for a mixture of hydrocarbons sample in a set-up with: a fixed bleed capillary, and an auxiliary pressure source.

Table S1 Comparison of peak volumes, along with $^1$D and $^2$D retention times for mixture of hydrocarbons sample obtained in two different set-ups. RSD - relative standard deviation.

| Compound name | Restrictor 1.3 m × 0.05 mm ID | | | Auxilliary pressure source F=9.5 mL/min | | | RSD (%) | | |
|---|---|---|---|---|---|---|---|---|---|
| | Blob Area | Retention time $^1$D | Retention time $^2$D | Blob Area | Retention time $^1$D | Retention time $^2$D | Blob Area | Retention time $^1$D | Retention time $^2$D |
| 1,2,4,5-Tetramethylbenzene | 229482.62 | 48.17 | 1.07 | 224540.20 | 48.25 | 1.08 | 1.54% | 0.12% | 0.66% |
| 1,2,4-Trimethylbenzene | 235003.62 | 38.00 | 1.08 | 232045.73 | 38.08 | 1.09 | 0.90% | 0.15% | 0.65% |
| 2-Ethyltoluene | 191523.82 | 36.67 | 1.12 | 188567.06 | 36.75 | 1.14 | 1.10% | 0.16% | 1.25% |
| Bicyclohexyl | 221992.26 | 62.92 | 0.81 | 218607.58 | 63.00 | 0.83 | 1.09% | 0.09% | 1.72% |
| Butylcyclohexane | 268622.98 | 42.17 | 0.58 | 263816.43 | 42.17 | 0.60 | 1.28% | 0.00% | 2.40% |
| C16:0 | 116483.29 | 98.08 | 0.67 | 111782.47 | 98.17 | 0.68 | 2.91% | 0.06% | 1.05% |
| C18:0 | 119529.09 | 107.58 | 0.64 | 115177.54 | 107.67 | 0.65 | 2.62% | 0.05% | 1.10% |
| C20:0 | 122657.94 | 116.33 | 0.61 | 118758.72 | 116.50 | 0.63 | 2.28% | 0.10% | 2.28% |
| C22:0 | 124097.57 | 124.42 | 0.59 | 120352.94 | 124.58 | 0.61 | 2.17% | 0.09% | 2.36% |
| Cyclododecane | 115908.34 | 64.42 | 0.83 | 113460.71 | 64.50 | 0.84 | 1.51% | 0.09% | 0.85% |
| Cyclooctane | 138678.11 | 32.08 | 0.79 | 137584.44 | 32.17 | 0.81 | 0.56% | 0.18% | 1.77% |
| Cyclopentadecane | 72268.74 | 82.92 | 0.82 | 70720.48 | 83.00 | 0.83 | 1.53% | 0.07% | 0.86% |
| Cyclopentylcyclohexane | 64125.53 | 54.42 | 0.80 | 62458.84 | 54.50 | 0.81 | 1.86% | 0.11% | 0.88% |
| Decaline | 211600.40 | 43.92 | 0.79 | 208083.89 | 44.00 | 0.80 | 1.18% | 0.13% | 0.89% |
| Ethylbenzene | 55230.02 | 26.08 | 1.09 | 54926.53 | 26.17 | 1.10 | 0.39% | 0.23% | 0.65% |
| Ethylcyclohexane | 57066.56 | 24.50 | 0.62 | 56604.78 | 24.58 | 0.64 | 0.57% | 0.24% | 2.24% |
| Ethylnaphthalene | 52927.69 | 67.58 | 1.64 | 50765.74 | 67.67 | 1.65 | 2.95% | 0.09% | 0.43% |
| Hexamethylbenzene | 49266.01 | 71.83 | 1.30 | 47830.98 | 71.92 | 1.31 | 2.09% | 0.08% | 0.54% |
| Hexylcyclohexane | 135597.20 | 58.08 | 0.56 | 132148.87 | 58.17 | 0.57 | 1.82% | 0.10% | 1.25% |
| Napththalene | 45005.42 | 52.58 | 1.85 | 43652.57 | 52.67 | 1.86 | 2.16% | 0.11% | 0.38% |
| nC10 | 149024.70 | 39.50 | 0.42 | 145370.19 | 39.58 | 0.44 | 1.76% | 0.15% | 3.29% |
| nC11 | 209756.12 | 47.75 | 0.40 | 204826.98 | 47.75 | 0.42 | 1.68% | 0.00% | 3.45% |
| nC12 | 289668.29 | 55.33 | 0.39 | 281919.28 | 55.42 | 0.40 | 1.92% | 0.11% | 1.79% |
| nC13 | 370605.15 | 62.58 | 0.38 | 361760.18 | 62.67 | 0.39 | 1.71% | 0.09% | 1.84% |
| nC14 | 432139.16 | 69.33 | 0.37 | 420867.23 | 69.33 | 0.38 | 1.87% | 0.00% | 1.89% |
| nC15 | 407073.17 | 75.58 | 0.36 | 397979.06 | 75.67 | 0.38 | 1.60% | 0.08% | 3.82% |
| nC16 | 338968.26 | 81.58 | 0.36 | 331372.59 | 81.67 | 0.38 | 1.60% | 0.07% | 3.82% |
| nC17 | 288284.28 | 87.17 | 0.35 | 281905.62 | 87.25 | 0.37 | 1.58% | 0.07% | 3.93% |
| nC18 | 240578.43 | 92.50 | 0.35 | 235987.04 | 92.50 | 0.36 | 1.36% | 0.00% | 1.99% |
| nC19 | 190076.64 | 97.58 | 0.35 | 185209.24 | 97.58 | 0.36 | 1.83% | 0.00% | 1.99% |
| nC20 | 143873.05 | 102.42 | 0.34 | 140735.18 | 102.50 | 0.36 | 1.56% | 0.06% | 4.04% |
| nC22 | 144324.95 | 111.58 | 0.34 | 140192.87 | 111.67 | 0.35 | 2.05% | 0.05% | 2.05% |
| nC24 | 120013.15 | 119.92 | 0.33 | 117037.36 | 120.08 | 0.35 | 1.78% | 0.10% | 4.16% |
| nC28 | 70175.86 | 134.83 | 0.32 | 68397.18 | 135.00 | 0.34 | 1.82% | 0.09% | 4.29% |
| nC30 | 46226.98 | 141.50 | 0.32 | 44935.59 | 141.67 | 0.33 | 2.00% | 0.08% | 2.18% |
| nC8 | 29154.37 | 21.75 | 0.44 | 28666.89 | 21.83 | 0.46 | 1.19% | 0.27% | 3.14% |
| nC9 | 82428.97 | 30.75 | 0.44 | 80500.26 | 30.83 | 0.45 | 1.67% | 0.19% | 1.59% |
| o-Xylene | 109386.86 | 28.92 | 1.15 | 108633.07 | 29.00 | 1.17 | 0.49% | 0.20% | 1.22% |
| Pentamethylbenzene | 181616.80 | 60.42 | 1.19 | 178361.86 | 60.50 | 1.20 | 1.28% | 0.10% | 0.59% |
| Pentylcyclohexane | 335277.03 | 50.50 | 0.57 | 325308.19 | 50.50 | 0.58 | 2.13% | 0.00% | 1.23% |
| Propylbenezne | 225088.28 | 34.50 | 1.03 | 222392.68 | 34.58 | 1.04 | 0.85% | 0.17% | 0.68% |
| Propylcyclohexane | 160289.05 | 33.33 | 0.60 | 158609.90 | 33.33 | 0.62 | 0.74% | 0.00% | 2.32% |
| Toluene | 25825.01 | 20.92 | 0.71 | 25851.71 | 21.00 | 0.72 | 0.07% | 0.28% | 0.99% |